\documentclass[aps,prl,twocolumn,superscriptaddress,showpacs]{revtex4}

\newcommand{\jgr}{J. Geophys. Res.}
\newcommand{\solphys}{Solar Phys.}

\begin{document}


\noindent 
{\bfseries Kliem and T\"or\"ok Reply:}
The Comment by Chen \cite{Chen2007} on our investigation of the torus
instability (TI) of arched magnetic flux ropes (hereafter KT
\cite{Kliem&T"or"ok2006}) states that he has previously solved a more
general equation than our Eq.~(1)---Eq.~(7) in
Ref.~\cite{Chen1989}---(including the effects of non-vanishing plasma
beta), using less restrictive assumptions (not assuming constant aspect
ratio $R/b$), so that ``the `torus instability' is a limiting case of
Ref.''~\cite{Chen1989}. In addition, it is claimed that \textsl{(i)} the
absence in KT of a term that introduces a spatial scale (the footpoint
distance $S_\mathrm{f}$ of an expanding rope whose ends are anchored in a
rigid surface) makes our model conflict with the presence of such a scale
in the observations of coronal mass ejections (CMEs) and \textsl{(ii)}
the assumption in KT's Eqs.~(8--10) that the ring current $I$ tends to be
conserved as a consequence of the rope's footpoint anchoring is
mathematically inconsistent with the ideal-MHD assumption that the
magnetic flux $\Psi$ enclosed by the rope is conserved.

The TI is a property of the tokamak equilibrium \cite{Shafranov1966} in
the low-beta limit, which is the relevant case in the coronal source
regions of CMEs. For $\beta<1$, the current ring can attain equilibrium
only in the presence of an external poloidal field $B_\mathrm{ex}$
\cite{Shafranov1966}. The instability occurs if $B_\mathrm{ex}$ decreases
sufficiently steeply with increasing major radius $R$, with a decay index
$n=-d\ln{B_\mathrm{ex}}/d\ln{R}\gtrsim3/2$ \cite{Bateman1978,
Kliem&T"or"ok2006}. In Ref.~\cite{Chen1989} the simplification
$B_\mathrm{ex}=0$ was adopted in the starting equations of both, the
linear and the nonlinear analysis (Eqs.~[7] and [32], respectively), thus
\textsl{excluding the TI}. (Moreover, since this requires $\beta>1$ for
equilibrium \cite{Shafranov1966}, the results in Ref.~\cite{Chen1989} are
largely irrelevant for the explanation of CMEs.)

Unlike Refs.~\cite{Chen1989, Chen1996, Chen&Krall2003}, we replaced the
integration of the force equation in the $b$ direction by the assumption
of self-similar expansion, $R/b=const$. This is required in order to
permit an analytical description (our Eqs.~[4--10], which have no
analogue in Refs.~\cite{Chen1989, Chen1996, Chen&Krall2003,
Chen&al2006}), and is also a very reasonable simplification, suggested by
the observations and justified by the facts that the hoop force depends
only logarithmically on $R/b$ and that our resulting description yields
qualitative and quantitative agreement with essential CME properties
\cite{Kliem&T"or"ok2006, T"or"ok&Kliem2007, Schrijver&al2007}. Chen
adopted this assumption recently as well \cite{Chen&al2006}, based on the
experiences made in fitting his model to CME observations. In integrating
the equation for $b(t)$ in Refs.~\cite{Chen1989, Chen1996,
Chen&Krall2003}, the energy equation was replaced by the polytropic
assumption, with the index $\gamma\approx1.2$ estimated from comparison
with observations; this is a level of approximation comparable to ours.
Our assumption $R/b=const$ obviously implies
$d^2b/dt^2\propto d^2R/dt^2$, not $d^2b/dt^2\simeq0$ as stated in the
Comment.

It is well known that the ``standard'' MHD equations (e.g.,
\cite{Bateman1978}) do not contain an intrinsic length scale. Dimensional
lengths enter by prescribing initial or boundary conditions at the
application stage, or by specifying the treatment such that these
conditions enter already explicitly at an intermediate stage. When our
scale-free Eq.~(4) is applied to describe CMEs, the footpoint anchoring
of the flux rope, which was not explicitly included in KT, introduces the
condition of nearly semicircular flux rope shape at TI onset (which is
supported by observations \cite{Vrsnak&al1991} and was suggested in
Ref.~\cite{Chen&Krall2003} as well); i.e., $R_0\approx S_\mathrm{f}/2$.
With the peak acceleration occurring in the range
$R\sim(1.5\mbox{--}2)R_0\approx(0.75\mbox{--}1)S_\mathrm{f}$ in the
practically most relevant range of $n$ included in Fig.~1 in KT, our
description reproduces the observational result of Fig.~1 in the Comment
to a very reasonable approximation, the more so if the spread in the
position of observed ejecta relative to the magnetic rope axis is taken
into account, and it does so in a manner that is more general than the
use of a problem-specific inductance in Refs.~\cite{Chen1996,
Chen&Krall2003, Chen&al2006} (which can easily be incorporated in our
theory). Contrary to a statement in the Comment, we did not suggest any
connection between $I(t)$ and the scale $S_f$.

$I$ and $\Psi$ can be simultaneously constant in a certain range of $R$.
This is obtained by using these two conditions joint with our general
ansatz $B_\mathrm{ex}(R)=\hat{B}R^{-n}$ in Eq.~(2) and elementary algebra
to solve it for $b/R$, given in Eq.~(10) and plotted in Fig.~3 of KT.
Using constant $I$, our Eq.~(8) follows exactly from our Eq.~(1). The
false conclusion in the Comment follows from the inappropriate
simplification $L\propto R$, neglecting the $\ln(R/b)$ dependence, which
removes $b$ from the equation.

Our estimate of the instability threshold $n_\mathrm{cr}$ for constant
$I$ (Eq.~[9]) includes an inconsistency because $R/b=const$ was used in
addition to constant $I$ and $\Psi$. However, \textsl{(i)} we have
expressed that Eqs.~(8--9) represent a limiting case, included to
demonstrate the \textsl{direction} of the effect of constant $I$,
\textsl{(ii)} this estimate does not play any further role in KT, and
\textsl{(iii)} the inconsistency of this approximation is made apparent
immediately following Eq.~(9) and Fig.~2 in KT.

\bigskip\bigskip 
B. Kliem$^1$ and T. T\"{o}r\"{o}k$^2$

\leftskip10pt
$^1$Kiepenheuer Institute, Freiburg, Germany \par
$^2$LESIA, Observatoire de Paris, CNRS, France

\leftskip0pt
\medskip\noindent 
Received: 23 October 2006, revised 2 Aug/21 Sep 2007 


\end{document}